\documentclass[sigconf]{acmart}

\usepackage{booktabs} 

\usepackage[utf8]{inputenc} 
\usepackage[T1]{fontenc}    
\usepackage{url}            
\usepackage{booktabs}       
\usepackage{amsfonts}       
\usepackage{nicefrac}       
\usepackage{microtype}      
\usepackage{graphicx} 
\usepackage{subcaption}
\usepackage{empheq}
\usepackage{xspace}
\usepackage{amsthm}
\usepackage{color}
\definecolor{mydarkblue}{rgb}{0,0.08,0.45}
\usepackage{enumitem}

\newcommand{\hide}[1]{}

\copyrightyear{2019}
\acmYear{2019} 
\setcopyright{iw3c2w3}
\acmConference[WWW '19]{Proceedings of the 2019 World Wide Web Conference}{May 13--17, 2019}{San Francisco, CA, USA}
\acmBooktitle{Proceedings of the 2019 World Wide Web Conference (WWW'19), May 13--17, 2019, San Francisco, CA, USA}
\acmPrice{}
\acmDOI{10.1145/3308558.3313747}
\acmISBN{978-1-4503-6674-8/19/05}

\usepackage{balance}

\usepackage{booktabs}       
\usepackage{amsfonts}       
\usepackage{microtype}      
\usepackage{xcolor}
\usepackage{url}
\usepackage{verbatim} 
\usepackage{graphicx}
\usepackage{caption} 
\usepackage{multirow}
 \usepackage{hyperref}       

\newcommand{\xhdr}[1]{{\noindent\bfseries #1}.}

\setlist[itemize]{leftmargin=*}
\usepackage{caption}

\title{Hierarchical Temporal Convolutional Networks for \\ Dynamic Recommender Systems}

\author{Jiaxuan You$^{1,3}$, Yichen Wang$^{2,3}$, Aditya Pal$^{3}$, Pong Eksombatchai$^{3}$, \\ Chuck Rosenberg$^{3}$, Jure Leskovec$^{1,3}$}

\affiliation{%
  \institution{Stanford University$^1$, Georgia Institute of Technology$^2$, Pinterest$^3$}
}
\email{{jiaxuan, jure}@cs.stanford.edu, yichen.wang@gatech.edu, {apal, pong, crosenberg}@pinterest.com}

\begin{document}

\begin{abstract}
Recommender systems that can learn from cross-session data to dynamically predict the next item a user will choose are crucial for online platforms. 
However, existing approaches often use out-of-the-box sequence models which are limited by speed and memory consumption, are often infeasible for production environments, and usually do not incorporate cross-session information, which is crucial for effective recommendations.
Here we propose Hierarchical Temporal Convolutional Networks (HierTCN), a hierarchical deep learning architecture that makes dynamic recommendations based on users' sequential multi-session interactions with items.
HierTCN is designed for web-scale systems with billions of items and hundreds of millions of users. It
consists of two levels of models: The high-level model uses Recurrent Neural Networks (RNN) to aggregate users’ evolving long-term interests across different sessions, while the low-level model is implemented with Temporal Convolutional Networks (TCN), utilizing both the long-term interests and the short-term interactions within sessions to predict the next interaction.
We conduct extensive experiments on a public XING dataset and a large-scale Pinterest dataset that contains 6 million users with 1.6 billion interactions. We show that HierTCN is 2.5x faster than RNN-based models and uses 90\% less data memory compared to TCN-based models. We further develop an effective data caching scheme and a queue-based mini-batch generator, enabling our model to be trained within 24 hours on a single GPU.
Our model consistently outperforms state-of-the-art dynamic recommendation methods, with up to 18\% improvement in recall and 10\% in mean reciprocal rank. 

\end{abstract}

\maketitle

{\fontsize{8pt}{8pt} \selectfont
\textbf{ACM Reference Format:}\\
Jiaxuan You, Yichen Wang, Aditya Pal, Pong Eksombatchai, Chuck Rosenberg, Jure Leskovec. 2019. Hierarchical Temporal Convolutional Networks for Dynamic Recommender Systems. In \textit{Proceedings of the 2019 World Wide Web Conf. (WWW'19), May 13--17, 2019, San Francisco, CA, USA.} ACM, NY, NY, USA, 11 pages. https://doi.org/10.1145/3308558.3313747 }

\section{Introduction}



\footnotetext[1]{Work done while at Pinterest.}
For many web applications, making item recommendations that match users' interests is of key importance.
Effective recommendations greatly improve users experience and retention, which leads to long-term increase in engagement.
In real-world scenarios, user's interests dynamically shift and evolve over time. While interests of a user across different sessions might depend on their long-term interests and are hence somewhat stable, the short-term in-session interests tend to evolve rapidly.
Therefore, an ideal recommender system should be able to capture both levels of user's dynamic interests and update those interests in real-time based on user's interactions.



The most direct data sources to build a recommender system are users' past sequences of interactions, which are abundant and well structured.
%
%
Currently, rule-based models are still widely used to make dynamic recommendations. Many works have shown that rule-based models, e.g., recommending the items with the largest number of interactions, is in fact a very strong baseline \cite{hidasi2015session,quadrana2017personalizing}.
Recent years have witnessed the power of sequence-based deep learning models in computer vision \cite{oord2016pixel}, natural language processing \cite{mikolov2010recurrent}, and graph structured data \cite{you2018graphrnn}.
In addition, recent works have applied sequence models for dynamic recommender systems \cite{hidasi2015session,wu2016personal,tang2018personalized} as well. 
These techniques usually involve using a specific sequence model, such as RNN or Convolutional Neural Networks (CNN), to encode users' past interactions into a latent feature space, which is then used for future predictions.

However, none of these approaches is suitable for recommender systems that scale to modern web-scale production environments with hundreds of millions of user, billions of items and tens of billions of interactions per day.
In particular, rule-based models fail to perform well in complex large-scale tasks.
Purely RNN-based or CNN-based approaches do not capture the hierarchical nature of in- and cross-session user interests.
In addition, RNN-based approaches are slow and difficult to train on massive data due to issues with gradient backpropagation.
CNN-based approaches have high memory consumption, and do not involve smooth and interpretable latent representations than can be reused for down-stream tasks.

Here we propose Hierarchical Temporal Convolutional Networks (HierTCN), a novel neural architecture for modeling users' sequential interactions, which enables real-time large-scale recommender systems. We design our model with modern web-scale recommender system production environment in mind; therefore, HierTCN is memory-efficient and fast to compute.
HierTCN consists of two levels of models to capture hierarchical levels of user interests. The high-level model uses RNN to capture users’ evolving long-term interests across sessions, while the low-level model is implemented with TCN, utilizing both a user's long-term interests and the short-term interactions within sessions to output a dynamic user embedding and make recommendations.
We apply HierTCN to a public dataset as well as a large private dataset with 1.7 billion training examples. We show that HierTCN has the following benefits over existing approaches:
\begin{enumerate}
    \item HierTCN has a significant performance improvement over existing deep learning models by about 30\% on a public XING dataset and 18\% on a private large-scale Pinterest dataset.
    \item Compared with RNN-based approaches, HierTCN is 2.5 times faster in terms of training time and allows for much easier gradient backpropagation.
    \item Compared with CNN-based approaches, HierTCN requires roughly 10\% data memory usage and allows for easy latent feature extraction.
\end{enumerate}
Based on HierTCN, we build a dynamic recommender system that scales to millions of users and billions of interactions. In terms of scale, it is at least 100 times larger than existing dynamic deep learning based recommender systems. The HierTCN-based recommender system features in the following components:
\begin{enumerate}
    \item A framework for joint modeling the dynamics of millions of users and items, which is not possible by existing approaches.
    \item An efficient offline training pipeline for HierTCN, which consists of efficient data caching and mini-batch generator.
    \item An efficient online inference pipeline that enables real-time recommendations using HierTCN.
\end{enumerate}

\section{Related Work}



\xhdr{Sequence models}
CNNs and RNNs are two important architectures for sequence modeling.
The sequential nature of RNN has made it the default choice for sequence modeling, and Long short-term Memory (LSTM) \cite{hochreiter1997long} and GRU \cite{chung2014empirical} are the two most popular RNN variants.
CNN has a long history of its application to sequence modeling as well \cite{lecun1995convolutional}. 
1D CNN with dilated convolutions have been shown to be powerful for audio data \cite{van2016wavenet}. The idea has been further developed by \cite{bai2018empirical} and is summarized as Temporal Convolutional Network (TCN), demonstrating its capability of modeling sequences in general. Our work here builds on this line of work but extends it by making it feasible for modeling hierarchical user interests and large-scale production environments.

\xhdr{Static recommender systems}
Recently, there has been a surge of interest in applying deep learning to recommendation systems. 
Several approaches treat user interactions as static information by ignoring the temporal dimension, and try to learn a static similarity matrix between users and items \cite{he2017neural,guo2017deepfm,hsieh2017collaborative,sedhain2015autorec}.
In contrast, our work explicitly considers the dynamics of user-item interactions through sequence modeling.

\xhdr{Dynamic recommender systems}
RNN is the most widely used architecture for dynamic recommender systems~\cite{wu2017recurrent,jannach2017recurrent,hidasi2015session,dai2016deep,quadrana2017personalizing}. 
Concretely,~\cite{wu2017recurrent} captures temporal aspects of user-item interactions and uses LSTM cell coupled with stationary factors to identify movie popularity fluctuations.~\cite{jannach2017recurrent} interpolates k-nearest-neighbor method with a session-based RNN~\cite{hidasi2015session} and demonstrates performance gains over static recommender systems. Recently,~\cite{dai2016deep,wang2016coevolutionary} combine point process models with the RNN-style state update functions to capture the co-evolution of user and item embeddings. However, these works only use one RNN network and do not explicitly consider multi-session settings.
~\cite{quadrana2017personalizing} is the most relevant work with our paper, where a hierarchical Gated Recurrent Units (GRU) framework is proposed. Our work differs in the way we design the hierarchy, the application of the TCN model, and many practical techniques that allow for deployment on real-world large-scale problems. Experimental results also show significant performance improvement of our model in both accuracy and speed. Furthermore, TCN-based sequence models remain largely unexplored for recommender systems, except for a very recent preprint~\cite{yuan2018simple}. In comparison, our work significantly alleviates the memory consumption issue of TCN which restricts its large-scale application, and explicitly models different levels of user interests.

\section{Proposed Model}

\subsection{Problem Setup}
\label{sec:problem}
We consider the problem of building a dynamic recommender system that adapts to users' preferences in real-time. For each user, we observe a sequence of items $\mathbf{c} = (\mathbf{c}_1,\mathbf{c}_2,...\mathbf{c}_n)$ that are impressed to a user, where $\mathbf{c}_t \in \mathbb{R}^{d \times m_t}$ are the embeddings of $m_t$ impressions at time $t$. 
The user interacts with some subset of the impressed items, which are denoted as $\mathbf{x} = (\mathbf{x}_1,\mathbf{x}_2,...\mathbf{x}_n)$, where $\mathbf{x}_t\in \mathbb{R}^d$ is the embedding of the interacted item at time step $t$, with $n$ being the total number of interactions for the user.
In our context, an interaction $\mathbf{x}_t$ refers to \textit{positive} actions such as clicking, saving, sharing, or buying an item, while impressions $\mathbf{c}_t$ refer to just viewing items.
In addition, we assume that interactions can be segmented into sessions using a function $q(\cdot)$, such that $i = q(t)$ indicates that the time step $t$ belongs to session $i$.
Our goal is to learn a function that predicts $\mathbf{x}_t$ from $\mathbf{c}_{\leq t}$ and $\mathbf{x}_{<t}$. 

\subsection{High-level Sketch of Our Approach}
Our solution is based on inferring user and item embeddings from the historical data ($\mathbf{x}, \mathbf{c})$. 
We note that typically users' interests vary significantly within sessions.
Hence, it is crucial to update user embeddings in near real-time in order to model their short-term interests accurately.
On the other-hand, item embeddings can be fairly stable and only require update after regular time-intervals and in-between these intervals they can be considered static for the sake of efficiency.
In practice, we  update item
embeddings daily, 
whereas user embeddings are updated in near real-time.
This choice is discussed in detail in Section \ref{sec:joint_model}.

The main goal of our recommender system is to predict the next interaction from a set of candidate items.
We formulate the problem using a function $f_{\theta}(\cdot)$ that infers the \textit{user embeddings} $\mathbf{u}_t$, then compute the similarity between $\mathbf{u}_t$ and the candidate item embeddings to produce an unnormalized conditional distribution $p(\mathbf{x}_t|\mathbf{x}_{<t})$. Specifically,
\begin{equation}
\label{eq:prob}
\begin{aligned}
    & p(\mathbf{x}_t|\mathbf{x}_{<t}) = S(\mathbf{x}_t, \mathbf{u}_t);
    & \mathbf{u}_t=f_{\theta}(\mathbf{x}_{<t})
\end{aligned}
\end{equation}
where $f_{\theta}(\cdot)$ is implemented as a deep neural network, and $S(\cdot)$ is a scoring function, e.g., a dot product, that measures the similarity between $\mathbf{x}_t$ and $\mathbf{u}_t$.
To finally make a recommendation, we evaluate $p(\mathbf{x}_t|\mathbf{x}_{<t})$ over a pool of candidate items and rank the items in descending order.
When the impression data $\mathbf{c}$ is known, the candidate pool consists of $m_t$ impressions at time step $t$; otherwise, the candidate pool could be all the items, a random subset of all the items, or a selection of items via some simple heuristics.
Next, we review the key ingredients that constitute $f_\theta(\cdot)$.

\subsection{Single-level Sequence Models}
We review the two recent state-of-the-art sequence models, which serve as the building blocks of our model.

\subsubsection{Recurrent Neural Networks}
\label{sec:rnn}
RNNs are naturally designed for modeling sequences. RNN maintains a hidden state vector that is updated with new inputs using the following update function:
\begin{equation}
    \mathbf{s}_t = \sigma(\mathbf{W}\mathbf{x}_t+\mathbf{U}\mathbf{s}_{t-1})
\end{equation}
where $\mathbf{x}_t$ is the input at time step $t$, $\mathbf{s}_t$ is the hidden state, and $\mathbf{W},\mathbf{U}$ are trainable parameters.
RNNs are known to suffer from vanishing gradient problem, and Gated Recurrent Unit (GRU) is a popular model that mitigates the issue \cite{chung2014empirical}. The update function of a GRU can be written as
\begin{equation}
\label{eq:gru}
\begin{aligned}
    & \mathbf{g}_t =  \sigma(\mathbf{W}^g\mathbf{x}_t+\mathbf{U}^g\mathbf{s}_{t-1})\\
    & \mathbf{r}_t =  \sigma(\mathbf{W}^r\mathbf{x}_t+\mathbf{U}^r\mathbf{s}_{t-1})\\
    & \mathbf{h}_t =  \tanh (\mathbf{W}^h\mathbf{x}_t+\mathbf{U}^h(\mathbf{s}_{t-1} \odot \mathbf{r_t}))\\
    & \mathbf{s}_t =  (1-\mathbf{g}_t) \odot \mathbf{h}_t + \mathbf{g}_t \odot \mathbf{s}_{t-1}
\end{aligned}
\end{equation}
where $\mathbf{x}_t$ is the input at time step $t$, $\mathbf{s}_t$ is the hidden state, $\mathbf{g}_t$ is the update gate, $\mathbf{r}_t$ is the reset gate, $\mathbf{h}_t$ is the candidate activation, $\sigma(\cdot)$ is the sigmoid function, and $\mathbf{W}^g,\mathbf{W}^r,\mathbf{W}^h,\mathbf{U}^g,\mathbf{U}^r,\mathbf{U}^h$ are trainable parameters. A Multilayer Perceptron (MLP) with ReLU activation is then used to output prediction $\mathbf{u}_{t+1}$:
\begin{equation}
    \mathbf{u}_{t+1} = \mathbf{W}^{(2)}\textsc{ReLU}(\mathbf{W}^{(1)}\mathbf{s}_t+\mathbf{b}^{(1)})+\mathbf{b}^{(2)}
\end{equation}
where $\mathbf{W}^{(1)},\mathbf{W}^{(2)},\mathbf{b}^{(1)},\mathbf{b}^{(2)}$ are trainable parameters.
Overall, GRU is a satisfying model for $f_{\theta}(\cdot)$, as $\mathbf{u}_{t+1}$ depends on all the previous interactions $\mathbf{x}_{\leq t}$.

\subsubsection{Temporal Convolutional Networks}
\label{sec:tcn}
TCN is a special type of 1D CNN, which is a natural way to encode information from a sequence \cite{bai2018empirical}.
A vanilla 1D convolutional layer can be written as
\begin{equation}
\label{eq:conv1d}
\begin{aligned}
    & F(\mathbf{x}_t) = (\mathbf{x}*\mathbf{f})(t) = \sum^{k-1}_{j=0}\mathbf{f}_j^T\mathbf{x}_{t-j}, t\geq k \\
    & \mathbf{u} = (F(\mathbf{x}_k),F(\mathbf{x}_{k+1}),...,F(\mathbf{x}_n))
\end{aligned}
\end{equation}
where $\mathbf{x}$ is the input sequence, $\mathbf{u}$ is the output sequence, and $\mathbf{f}\in \mathbb{R}^{k \times d}$ is a convolution filter with size $k$. A 1D CNN is then constructed by stacking several vanilla 1D convolutional layers. However, 1D CNN is restricted by its shrinking output size and limited receptive fields when being applied to model sequences, while TCN features with two techniques that solve these problems namely causal convolutions and dilated convolutions.


\xhdr{Causal convolutions}
As is shown in Equation \ref{eq:conv1d}, a vanilla 1D convolutional layer takes as input a length $n$ sequence and outputs a length $n-k+1$ sequence.
The output can shrink further if more such layers are stacked together.
This property can be problematic in our domain, as we want our model to make predictions at every time step and make updates in real-time. 
A causal convolutional layer solves the problem by concatenating a length $k-1$ zero padding at the beginning of the input sequence.
Furthermore, it ensures that there is no information leakage from the future into the past, which is crucial when predicting future interactions. Concretely,
\begin{equation}
\label{eq:conv1d_causal}
\begin{aligned}
    & F(\mathbf{x}_t) = (\mathbf{x}*\mathbf{f})(t) = \sum^{k-1}_{j=0}\mathbf{f}_j^T\mathbf{x}_{t-j} & \mathbf{x}_{\leq 0} := \mathbf{0} \\
    & \mathbf{u} = (F(\mathbf{x}_1),F(\mathbf{x}_{2}),...,F(\mathbf{x}_n))
\end{aligned}
\end{equation}
This formulation ensures that the output sequence $\mathbf{u}$ is well-defined over each time step, and prediction $\mathbf{u}_t$ only depends on input $\mathbf{x}_{\leq t}$.

\xhdr{Dilated convolutions}
Another issue with vanilla 1D CNN is that it has a receptive field linear to the number of layers, which in our case is undesirable since we aim to model long-term dependencies.
Dilated convolution is a technique that allows for receptive fields exponential to the number of layers.
Specifically, when combining with causal convolution, the $r^{\text{th}}$ level dilated convolutional layer can be written as
\begin{equation}
\label{eq:conv1d_dilated}
\begin{aligned}
    & F(\mathbf{x}_t) = (\mathbf{x}*_{l_r}\mathbf{f})(t) = \sum^{k-1}_{j=0}\mathbf{f}_j^T\mathbf{x}_{t-{l_r}\cdot j} & \mathbf{x}_{\leq 0} := \mathbf{0}\\
    & \mathbf{u} = (F(\mathbf{x}_1),F(\mathbf{x}_{2}),...,F(\mathbf{x}_n))
\end{aligned}
\end{equation}
where $l_r$ is the dilation factor which can be set as ${(k-1)}^{r-1}$ to achieve exponentially large receptive field. We refer to this formulation in Equation \ref{eq:conv1d_dilated} as temporal convolutional layer.

A Temporal Convolutional Network (TCN) is then constructed by stacking multiple temporal convolutional layers. To facilitate training a deep TCN, a common practice is to organize temporal convolutional layers into blocks, and add residual connections \cite{he2016deep} between blocks. By setting a proper size of filter and number of layers, $\mathbf{u}_{t+1}$ can depend on the full historical interactions $\mathbf{x}_{\leq t}$.

    



\begin{figure*}
    \centering
    \includegraphics[width=0.7\linewidth]{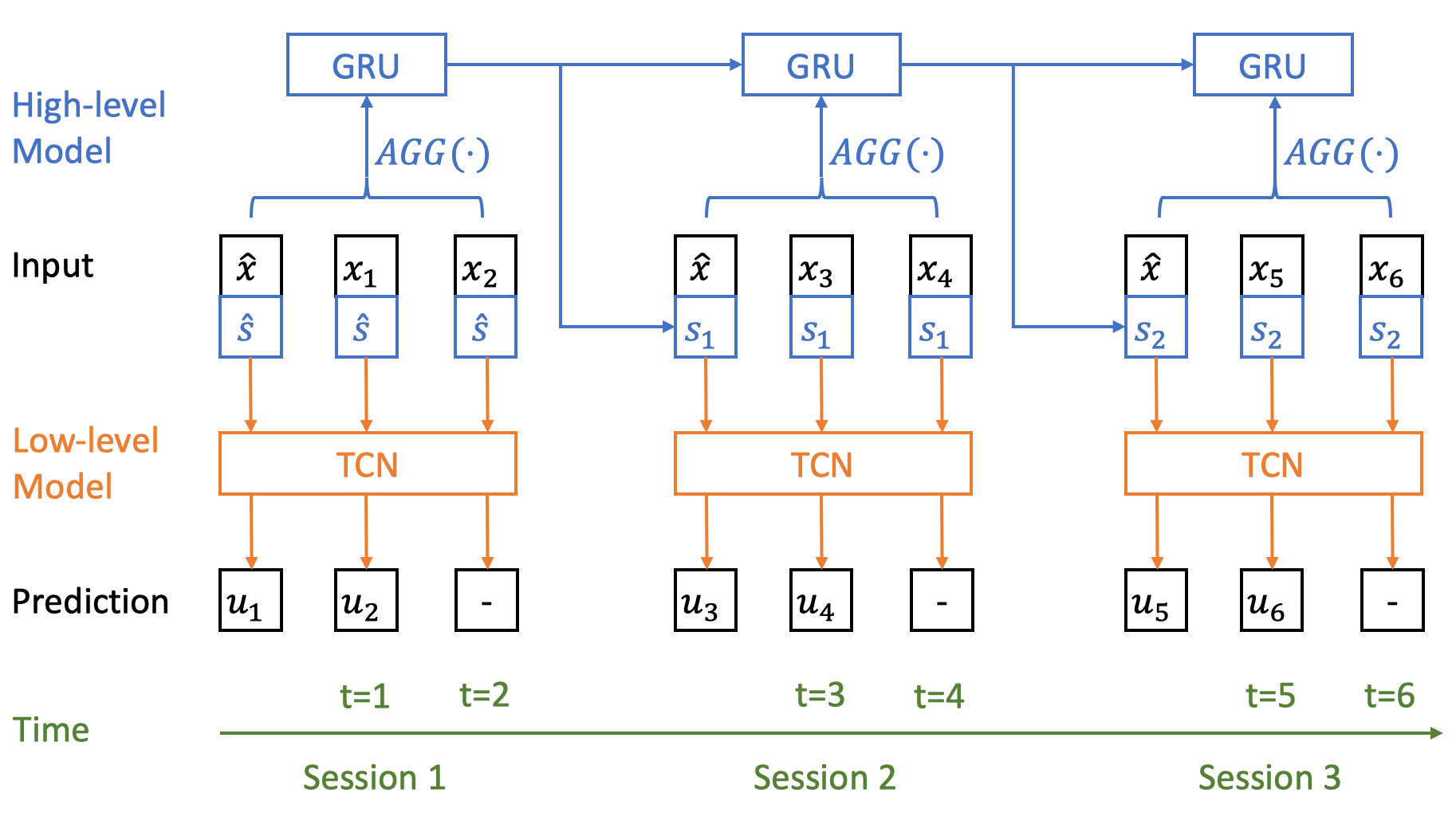}
    \vspace{-4mm}
    \caption{Visualization of HierTCN architecture. HierTCN generates predictions $\mathbf{u}_1,\mathbf{u}_2,...\mathbf{u}_n$ based on a sequence of interactions $\mathbf{x}_1,\mathbf{x}_2,...\mathbf{x}_n$. The high-level model (blue) is implemented with GRU which is updated by an aggregation of each session of interactions using function $\textsc{AGG}(\cdot)$. The low-level model (orange) uses TCN to predict user embeddings at each time step, based on a user's past interactions within the session and the hidden state $\mathbf{s}_i$ of the high-level model. $\hat{\mathbf{x}}$, $\hat{\mathbf{s}}$ are the default start tokens, which are used to produce the first output for the low-level model and the high-level model, respectively.}
    \label{fig:HierTCN}
    \vspace{-3mm}
\end{figure*}

\subsection{Design Choices of Hierarchical Sequence Models}
\label{sec:consideration}
It is reasonable to consider using one of the above single-level sequence models to predict users' future interactions based on their past interactions.
However, this formulation omits the session information which implies a hierarchy of user interests. 
Users typically have long-term interests that span across multiple sessions.
These interests need to be updated after the end of each session, i.e. in the order of few hours to days.
On the other hand, users' short-term interests are reflected via the interactions within each session, which must be updated in the order of a few seconds.
In principle, we could include the session information as a feature to the single-level sequence models, enabling them to learn the short- and long-term interests.
However, our experiments show that this approach does not work well in practice.
The issue is that such an approach does not have the inductive bias over the hierarchy of user interests and thus fails to generalize to unseen cases.

Rather than directly applying single-level sequence models, we impose the inductive biases over users' hierarchical interests via incorporating the hierarchical structure in the design of the neural network architecture.
Specifically, our hierarchical model has a low-level component that learns from interactions within a session, and a high-level component that carries over information across sessions.
The low-level component outputs a user embedding $\mathbf{u}_t$ using a neural network $f_{\omega}(\cdot)$, which can be written as:
\begin{equation}
    \mathbf{u}_t = f_{\omega}((\mathbf{x}_i|\forall q(i)=q(t), i<t),\mathbf{s}_{q(t)})
\end{equation}
where $(\mathbf{x}_i|\forall q(i)=q(t), i<t)$ is the sequence of within-session item embeddings before time step $t$, and $s_{q(t)}$ represents a user's long-term interest which is modeled by a high-level component:
\begin{equation}
    \mathbf{s}_{q(t)} = f_{\phi}((\mathbf{x}_i|\forall q(i)<q(t)))
\end{equation}
where $f_{\phi}(\cdot)$ is the high-level model that summarizes the sequence of interactions before the session $q(t)$.
In the remainder of this section, we will discuss different ways of designing hierarchical sequence models over multi-session data.


\subsubsection{Choices of sequence models}
\label{sec:choice_seq}
The low-level model focuses on predicting the next interaction, based on both within-session interactions and the high-level information. The task is similar to making sequential prediction without considering the session structure, but over a much shorter sequence compared with taking all historical interactions of a user as input. Although RNN is generally much slower than TCN since the computation cannot be paralleled, both RNN and TCN can be chosen to represent $f_{\omega}(\cdot)$.

For the high-level model, the goal is to model a user's long-term interests.
For this task, RNN has a natural advantage, because it maintains a hidden representation over different time steps, which can be naturally interpreted as a representation of the user's long-term interests.
In contrast, TCN calculates the output directly through layers of convolution.
Since there is no hidden state being maintained, the output is less smooth and interpretable, which is undesirable, especially when we would like to model a relatively stable long-term interests for all users.
Moreover, TCN requires taking the raw input sequences as input. For the high-level model, this requirement corresponds to keeping track of users' entire history of interactions whenever inference is needed, resulting in high data memory usage, which is undesirable for large-scale implementation. Thus, we conclude that RNN is more suitable for representing $f_{\phi}(\cdot)$.

\subsubsection{Choices of combining different levels of models}
\label{sec:choice_combine}
There are two problems to consider when trying to combine the low-level and the high-level model.
The first problem is how to condition the low-level prediction on the high-level information, to which there are two approaches.
The first approach is to use the high-level representation to initialize the low-level model; the second approach is to enforce the high-level representation propagated into the low-level model at each time step. These two approaches are first explored in \cite{quadrana2017personalizing}, and we refer to them as \textit{Init connection} and \textit{Full connection}, respectively.

The second problem is to update the high-level model with low-level information.
Suppose a user has $n$ total interactions over $m$ sessions, and function $\textsc{AGG}(\cdot)$ aggregates the embeddings within a session whose output is then used to update the high-level model.
\cite{quadrana2017personalizing} explores using the low-level model as the aggregation function, where the final hidden state of a RNN is used to update the high-level model.
However, this approach complicates the path of gradient backpropagation, as the longest backpropagation path has a length of $n$. In addition, the method couples the training of the low-level model and the high-level model; consequently, the biased low-level model at the start of training can negatively affect the training of the high-level model, resulting in a slow convergence. 
In contrast, we propose to use a simpler $\textsc{AGG}(\cdot)$ function, e.g., the $\textsc{Mean}(\cdot)$ pooling function, by assuming that the ordering of within-session interactions carries little information.
Using a $\textsc{Mean}(\cdot)$ pooling function only results in a length-$m$ backpropagation path and decouples the training of the low-level model and the high-level model. We empirically find that this approach converges much faster and provides much better prediction performance.

\subsection{HierTCN}
\label{sec:HierTCN}
Based on the discussions in Section \ref{sec:consideration}, we design HierTCN, an efficient and scalable hierarchical sequence model. 
\subsubsection{Model design}
HierTCN consists of a  high-level model implemented with GRU and a low-level model implemented with TCN.
The long-term user interests are represented via GRU's hidden state. The hidden state is updated by an aggregation over item embeddings within each session, using the $\textsc{Mean}(\cdot)$ pooling function. Specifically, after all the interactions in a session $q(t)$ have been observed, user's long-term interests, represented as the GRU hidden state $\mathbf{s}_{q(t)}$, are updated using the following equation:
\begin{equation}
    \mathbf{s}_{q(t)} = \textsc{GRU}(\mathbf{s}_{q(t)-1},\textsc{Mean}((\mathbf{x}_i|\forall q(i)=q(t))))
\end{equation}
where $q(t)$ refers to the session that contains time step $t$, $(\mathbf{x}_i|\forall q(i)=q(t))$ is the sequence of item embeddings in the session $q(t)$, and $\textsc{GRU}(\cdot)$ is a GRU described in Section \ref{sec:rnn}.

The low-level TCN model represents short-term interests of the users and predicts their next interactions within the session. We use the \textit{Full connection} (Section \ref{sec:choice_combine}), i.e., include $\mathbf{s}_{q(t)}$ as the input to the TCN at each time step. 
TCN outputs a prediction $\mathbf{u}_{t}$, which can be interpreted as a dynamic user embedding.
Concretely,
\begin{equation}
    \begin{aligned}
    & \mathbf{u}_t = \textsc{TCN}((\textsc{concat}(\mathbf{x}_i,\mathbf{s}_{q(t)-1})|\forall q(i)=q(t), i<t))
\end{aligned}
\end{equation}
where $\textsc{concat}(\cdot)$ concatenates an input item embedding with the corresponding high-level hidden state,  $(\mathbf{x}_i|\forall q(i)=q(t), i<t)$ is the sequence of within-session item embeddings before time step $t$, $\mathbf{s}_{q(t)}$ is passed from the high-level model and propagates its gradients into the high-level model when being trained, and $\textsc{TCN}(\cdot)$ is a TCN described in Section \ref{sec:tcn}.

To output a final recommendation, HierTCN uses dot product between user and item embeddings as the scoring function to compute the unnormalized distribution $p(\mathbf{x}_t|\mathbf{x}_{<t})$
\begin{equation}
    p(\mathbf{x}_t|\mathbf{x}_{<t}) = \mathbf{x}_t^T \mathbf{u}_t
\end{equation}
We then compute $p(\mathbf{x}_t|\mathbf{x}_{<t})$ over the items in a candidate pool and rank the items in descending order, and output the top $k$ items as the final recommendations.
The overall architecture of HierTCN is visualized in Figure \ref{fig:HierTCN}.


\subsubsection{Objective function}
\label{sec:objective_function}

In this section, we discuss the possible objective functions than can be used to train HierTCN when being applied to recommender systems. 
The performances of differ objective functions are compared in Section \ref{sec:objective_result}
When items can be represented as real-value vectors, the following objective functions are commonly used.
\begin{itemize}

\item \textbf{L2 loss} 
The simplest objective function would be minimizing the L2 distance between the user embedding and the interacted item embedding:
\begin{equation}
    \min || \mathbf{x}_t - \mathbf{u}_t ||_2
\end{equation}

\item \textbf{Noise Contrastive Estimation (NCE)}
This objective is first proposed by \cite{gutmann2012noise}, and is popularized by the Word2Vec paper \cite{mikolov2013distributed}. This objective function makes use of the negative samples. Specifically, the objective function encourages the user embedding to be similar to the positive item embedding, while enforcing the user embedding to be different from the negative items' embeddings. NCE employs the following formulation:
\begin{equation}
    \min -\log(\sigma(\mathbf{x}_t^T \mathbf{u}_t)) - \sum_i -\log(\sigma(-\mathbf{c}_{ti}^T \mathbf{u}_t))
\end{equation}

\item \textbf{Bayesian personalized ranking (BPR)}
\cite{rendle2009bpr} proposes using the following BPR objective function, which can be interpreted as maximizing the posterior estimator derived from a Bayesian analysis of the item recommendation problem:
\begin{equation}
    \min - \sum_i \log(\sigma(\mathbf{c}_{ti}^T \mathbf{u}_t - \mathbf{x}_t^T\mathbf{u}_t))
\end{equation}

\item \textbf{Hinge loss}
Hinge loss is based on the idea of max margin learning \cite{suykens1999least}, which has the following form:
\begin{equation}
    \min \sum_i \max \Big\{ 0, \delta + \mathbf{c}_{ti}^T \mathbf{u}_t - \mathbf{x}_t^T \mathbf{u}_t \Big \}
\end{equation}

\end{itemize}

When items do not come with features and thus are represented as one-hot vectors, the following cross entropy loss can be used:
\begin{itemize}
    \item \textbf{Cross entropy loss}
    When $\mathbf{x}_t$ are one-hot vectors, cross entropy loss can be written as
\begin{equation}
    \min - \mathbf{x}_t^T \log(\mathbf{u}_t)
\end{equation}

\end{itemize}

\subsubsection{Regularization and Normalization}
There are many techniques to facilitate model convergence and prevent overfitting, and we adopt the dropout \cite{srivastava2014dropout} and batch normalization \cite{ioffe2015batch} in the HierTCN model for both TCN and GRU modules.
These techniques are not widely adopted for sequence models, and we provide a thorough discussion in this section.

\xhdr{Dropout}
Dropout can be easily implemented for TCN by randomly dropping activations for the subsequent layers.
For RNN, normally dropout is applied at the input, the output and the state transition function \cite{abadi2016tensorflow}. 
However, the benefits of applying dropout for state transition function are in fact doubtful because early state information can be wiped out by an exponential factor over the sequence length $n$. For recommender systems where users usually generate long interaction sequences, this approach is especially undesirable.

We use the following approach~\cite{semeniuta2016recurrent} to applying dropout for GRU transition. Specifically, dropout is only applied to the candidate activation $\mathbf{h}_t$, and only the last equation in Equation \ref{eq:gru} is modified into $\mathbf{s}_t = (1-\mathbf{g}_t) \odot D(\mathbf{h}_t) + \mathbf{g}_t \odot \mathbf{s}_{t-1}$, where $D(\cdot)$ is the dropout function. This technique does not wipe out information from the last time step, while still adding randomness to the state transition to prevent overfitting.

\xhdr{Batch normalization}
When carrying out batch normalization, we find it necessary to keep track of the average statistics over each time step independently \cite{cooijmans2016recurrent}. 
For dynamic recommender systems, since different users can have diverse sequence lengths, a common technique is to zero pad all the sequences to the same length. When updating the average statistics, it is important to mask out zero padded values in the batch normalization layer, in order to prevent it from using these meaningless values to update the statistics.


\subsection{HierTCN for Real-world Recommender Systems}

In this section, we discuss how to apply HierTCN to build a real-world dynamic recommender system.
Specifically, we propose a general framework that allows for a joint update of user and item embeddings, and discuss some key techniques to ensure efficient model training and inference.


\subsubsection{Joint modeling of user and item embeddings}
\label{sec:joint_model}

When applying HierTCN for recommender systems, we propose two separate update mechanism: (1) User interests change rapidly and thus their embeddings get updated in real-time. (2) Item information changes more slowly and thus we update item embeddings on a daily schedule. This choice turns out to be very important in practice. Existing approaches that learn latent variables over all the interactions in a global chronological order between all the users and items \cite{dai2016deep} cannot scale to billions of interactions that are observed in a real-world recommender system. Such models become computationally prohibitive and infeasible for large datasets.

We side-step this problem by designing a two-phase update scheme. In the first phase, item embeddings are generated. The second phase considers the item embeddings fixed and only generates user embeddings.


\xhdr{Update of item embeddings}
In the first phase, we first build an item graph based on user-item interactions, where each node represents an item and an edge is built if two items are interacted by the same user within a short period of time.
Then, we compute the node embedding for each item, which integrates both visual and textual features as well as the structural features of the item graph using graph convolutional neural networks (GCN) based on localized graph convolutional layers \cite{ying2018graph}, whose computation can be paralleled on a distributed system. 
Specifically, each node is initialized with the concatenated visual and textual embeddings of the item. We then apply 2 layers of localized graph convolutional layers to compute the node embedding for a node $u$, where the $l^{\text{th}}$ layer can be written as
\begin{equation}
\begin{aligned}
    & \mathbf{n}_u^{(l)} = \textsc{AGG}(\textsc{ReLU}(\mathbf{Q}^{(l)}\mathbf{z}_v^{(l)}+\mathbf{q}^{(l)}|v\in N(u))) \\
    & \mathbf{z}_u^{(l+1)} = \textsc{ReLU}(\mathbf{W}^{(l)}\textsc{concat}(\mathbf{z}_u^{(l)}, \mathbf{n}_u^{(l)}))
\end{aligned}
\end{equation}
where $\mathbf{z}_u^{(l)}$ is the $l^{\text{th}}$ level node embedding for node $u$, $N(u)$ is the local neighborhood of $u$, $\textsc{AGG}(\cdot)$ is an order invariant aggregation function such as $\textsc{Mean}(\cdot)$ pooling, and $\mathbf{Q}^{(l)}, \mathbf{q}^{(l)}, \mathbf{W}^{(l)}$ are trainable parameters.
After gathering new user-item interactions, we update the item graph and recompute the node embeddings for each item.

\xhdr{Update of user embeddings}
In the second phase, we keep the item embeddings fixed and train the HierTCN model using the past interactions of users.
Once the model is trained, any new within-session or across-session activity leads to a relatively fast update of the HierTCN model.
Model can then be used in parallel to compute user embeddings in real time.


\xhdr{Training user and item models}
The GCN model is trained by minimizing the following NCE-based objective function
\begin{equation}
    -\log(\sigma(\mathbf{z}_u^T \mathbf{z}_v)) - C \mathbb{E}_{v_n\sim P_n(u)}[\log(\sigma(-\mathbf{z}_u^T\mathbf{z}_{v_n}))], \forall v \in N(u)
\end{equation}
where $P_n(u)$ is a negative sampling distribution for node $u$ and $C$ is the number of negative samples, while HierTCN is trained with the hinge loss described in Section \ref{sec:objective_function}.
GCN model is trained with about 7.5 billion samples and the HierTCN model is trained with about 1.7 billion samples, more details of training the GCN model are described in \cite{ying2018graph} and HierTCN training techniques are discussed in \ref{sec:offline_training}. 

Overall, this two-phase approach enables the dynamic modeling of both user and item embeddings, which can be computationally cost-prohibitive otherwise.
We conduct extensive offline evaluation on a subset of 6 million Pinterest users and present our results in Section~\ref{sec:experiments}.

\begin{table}
  \caption{Statistics of the datasets (mean$|$std. deviation). Note that XING dataset does not have the impression data.}
  \label{tb:data}
  \begin{tabular}{ccc}
    \toprule
  Dataset & XING & Pinterest  \\
    \midrule
    Users & 65,347& 5,923,659 \\
    Items & 20,778 &  74,202,787 \\
    Interactions & 1,450,300& 56,050,857 \\
    Impressions &-& 1,685,877,684 \\
    Sessions & 535,747 & 23,354,523  \\
    Impressions per event &-& 9.8|8.6 \\
    Impressions per session &-& 25.8|51.1 \\
    Events per session & 2.5|3.8 & 2.4|3.0 \\
    Sessions per user & 8.2|7.0 & 3.9|4.7 \\
    Events per user & 22.2|20.8 & 9.7|13.4 \\
    Impression per user &-& 284.6 | 252.0 \\
  \bottomrule
\end{tabular}
\vspace{-2mm}
\end{table}

\subsubsection{Efficient offline training of HierTCN}
\label{sec:offline_training}

HierTCN is trained offline with about 6 million users and 1.7 billion training examples (Table \ref{tb:data}).
To conduct the training, we design an efficient offline training pipeline for HierTCN, which includes the following key components.


\xhdr{Item embedding cache}
The 
item embeddings are represented as 512 dimensional float vectors, thus directly storing 1.7 billion interactions requires tens of terabytes of storage and even loading all the data to memory takes a day.
To speed up the data pipeline, we first assign a universal ID to each of the 74 million items, and only keep item IDs when saving the user interaction data. Then we employ Linux huge page table \cite{lameter2006local} to cache all the item embeddings in a fixed chunk of memory. 
Finally, a sequence of item embeddings can be fetched by a direct look-up in the page table with a negligible query time.
This approach enables us to load the item embeddings once; then, any subsequent models can share the embedding cache without the need to reload/rebuild the cache. In practice, this approach can reduce the data loading time from a day to a minute when launching a sequence model.
Note that in the online deployment setting, the item embedding cache is rebuilt daily.

\xhdr{Queue-based mini-batch generator}
To train a HierTCN, a mini-batch of training data should consist of the same number of sessions.
In practice, the number of sessions that a user has can range from 1 to several hundreds; thus, the common approach of doing zero padding will cause huge amount of unnecessary data memory consumption.
To address the problem, we design a queue-based mini-batch generator to significantly reduce memory consumption and speed up training, based on the data loading scheme proposed in \cite{quadrana2017personalizing}.
The mini-batch generator first initializes $B$ data queues where $B$ equals the batch size. Then an enqueue process loads a new user's sequence data and parses them by sessions, and enqueues the sessions of data along with the user ID to the queue with the least number of sessions. 
At the same time, a dequeue process takes a number of sessions from each queue, with each session zero-padded to the same length, then concatenates the sessions into batches and feeds a batch of data into HierTCN.
During computation, HierTCN model constantly checks whether the user ID changes after a session; if so, the model will reset the hidden state of the high-level GRU to ensure each new user's state is properly initialized. 
The proposed mini-batch generator also naturally conducts the truncated backpropagation technique \cite{williams1990efficient}, a key technique that ensures stable training of RNN over long sequences.

\subsubsection{Real-time inference}
For real-time inference, the recommender system consists of a mini-batch generator that produces mini-batches and a trained HierTCN model that consumes the mini-batches.
The mini-batch generator can be paralleled by creating multiple enqueue and dequeue processes, while the computation of HierTCN model can be paralleled by deploying multiple copies of the trained model, in order to achieve fast real-time computation. Two of the key techniques are discussed below.


\xhdr{User hidden state cache}
A real-time dynamic recommender system requires efficient update of users' hidden states.
Similar to item embedding cache, we cache all the users' hidden states into the memory, and maintain a dictionary that maps a user to the corresponding hidden state. Whenever a user starts a session, the system will read the hidden state of the user.
After a session is closed, the hidden state of a user is updated with the high-level GRU, and is written to the original memory address.

\xhdr{Online queue-based mini-batch generator}
We extend the offline mini-batch generator into the online settings.
Specifically, after a user has interacted with an item, an enqueue process will fetch a session of data that contains the user's past interactions within the session. In addition, the enqueue process reads the cached hidden state of the user and concatenates the hidden state vector to all the interactions within the session. Finally, the process enqueues the session of data.
At the same time, a dequeue process dequeues a batch of sessions with each session zero-padded to the same length and feeds them to HierTCN model. The process also retrieves the recommendation results and sends the results to the web API, while writing the updated hidden states to the cache.

\section{Experiments}



\label{sec:experiments}


\subsection{Datasets}
\label{sec:dataset}
We use a public XING dataset and a large-scale private Pinterest dataset. The statistics of the datasets are summarized in Table \ref{tb:data}.
The aim of experimenting on a small-scale public dataset is to demonstrate that HierTCN can achieve transferable and robust performance, while rigorous and extensive evaluations are done over the large-scale Pinterest dataset.

The public XING dataset is extracted from XING Recsys Challenge 2016 dataset \cite{2016recsys}, with about 11 thousand users and 500 thousand interactions. The items do not come with features thus we represent each item as a one-hot vector. The dataset also does not have session information, thus we manually partition the interactions using a 30-minute idle threshold. 
Following the prior work \cite{quadrana2017personalizing}, we remove interactions with type ``delete'' and do not consider the interaction types in the data. We remove items with less than 50 interactions and users with less than 10 or more than 1000 interactions.

The private Pinterest dataset is an internal dataset from Pinterest, with 6 million users, 56 million interactions and 1.7 billion impressions over 3 months. We clean the dataset by removing users with less than 10 or more than 1000 interactions. We represent items using a Pinterest internal item graph based on user-item interactions, with 3 billion nodes and 18 billion edges. Specifically, each item is represented as a node, and we connect two items if an user has interacted with both of them within a very short period of time. Each item is initialized with 4096 dimensional visual features, extracted from the 6-th fully connected layer of an image classification network using VGG-16 architecture \cite{simonyan2014very}, and 256 dimensional textual annotation features trained with Word2Vec \cite{mikolov2013distributed}. A 2-layer GraphSage model \cite{hamilton2017inductive} is then trained over the item graph using hinge loss function to differentiate positive and negative item samples. Finally, each item is represented by a 512 dimensional node embedding vector. More details about learning the item embeddings are discussed in \cite{ying2018graph}.

\begin{table*}
  \caption{Performance of cold-start recommendation on the XING dataset and the large-scale Pinterest dataset.
  }
  \vspace{-1mm}
  \label{tb:cold}
  \begin{tabular}{cc|c|cc|cc|cc}
    \toprule
  Dataset & Metric & \textbf{HierTCN} & HierGRU  & HRNN \cite{quadrana2017personalizing} & TCN & GRU & MV & MaxItem\\
    \midrule
     XING & Recall@10 (Higher is better) & \textbf{0.139}& 0.105  & 0.113& 0.108  & 0.124 & 0.107 & -\\
            & MRR (Higher is better) & \textbf{0.071} & 0.037  & 0.040 & 0.038 & 0.043 & - & - \\
            & MRP (Lower is better) & \textbf{0.121} & 0.161  & 0.149 & 0.198 & 0.141 & - & - \\
    \midrule
    
        Pinterest & Recall@1 & \textbf{0.206} & 0.184& 0.174  & 0.194 & 0.195 & 0.1658& 0.1487 \\
                & Recall@5 & \textbf{0.663} & 0.639 & 0.619 & 0.653 & 0.655 & 0.6120& 0.5843 \\
                & Recall@10 & \textbf{0.855} & 0.839 & 0.828 & 0.850 & 0.852 & 0.8173& 0.7921 \\
                & MRR & \textbf{0.402} & 0.380 & 0.366 & 0.391 & 0.392 & 0.3593& 0.3400 \\
                 & MRP & \textbf{0.304} & 0.326 & 0.338 & 0.313 & 0.311 & 0.3484& 0.3724 \\
  \bottomrule
\end{tabular}
\vspace{-1mm}
\end{table*}

\begin{table*}
  \caption{Performance of warm-start recommendation on the large-scale Pinterest dataset.} 
  \vspace{-1mm}
  \label{tb:pinterest_warm}
  \begin{tabular}{cc|c|c|c|cc}
    \toprule
  Dataset & Metric &\textbf{HierTCN} & HRNN \cite{quadrana2017personalizing} & TCN & MV & MaxItem\\
    \midrule
    Pinterest & Recall@1 &\textbf{0.202} & 0.188& 0.183 & 0.174& 0.155 \\
    & Recall@5 & \textbf{0.667} & 0.650 & 0.641  & 0.631& 0.600 \\
    & Recall@10 & \textbf{0.855} & 0.848 & 0.838 & 0.830& 0.801 \\
    & MRR & \textbf{0.399} & 0.386& 0.379 &  0.370& 0.348 \\
    & MRP & \textbf{0.313} & 0.324& 0.336 &  0.346& 0.374 \\
  \bottomrule
\end{tabular}
\vspace{-1mm}
\end{table*}


\subsection{Experimental Setup}

\subsubsection{Experimental setting}
We consider two types of experiments.
The first is \textit{cold-start recommendations}. We split all the data by user, and select 80\% of the users to train the model, 10\% to tune the hyper-parameters of the model and test on the remaining 10\% users. 
The second is \textit{warm-start recommendations}. We select a fixed set of users, then train the model on the first two months of data and test on data in the following month. All the hyper-parameters remain the same with the cold-start recommendation settings.

All the deep learning based models are trained with Adam optimizer, with learning rate 0.001 and batch size 32. We stop training the model when the validation error plateaus. We find that weight normalization \cite{salimans2016weight} does not help with the tasks we are experimenting, while batch normalization \cite{ioffe2015batch} and dropout \cite{srivastava2014dropout} is helpful for the smaller XING dataset. For the large-scale Pinterest dataset, the vanilla models without regularization perform well.

\subsubsection{Evaluation metrics}
Our primary goal is to predict the user interaction in the next time step, which is evaluated by ranking the ground truth interaction against a pool of candidates at the given time step. For the XING dataset, the candidate pool is the set of all the items, while for the Pinterest dataset, the candidate pool is the impression data at the evaluated time step, which is given by a separate production system and cannot be altered by the model.

For models that can directly produce the ranking of an item (e.g., Maximum item similarity (MaxItem)), we directly use the predicted ranking for evaluation.
For models that can output specific user embeddings, we compute the probability $p(\mathbf{x}_t|\mathbf{x}_{<t})$ over the candidate pool and rank the probability in descending order. We repeat the same process over all the interactions for all the users in the test data, and report various ranking statistics that are listed below.

\begin{itemize}
\item {\bf Recall@K}. Recall@K reports the proportion of times that the ground truth interacted item is ranked within the top K list of recommendations. Higher Recall@K is better.

\item {\bf Mean Reciprocal Rank (MRR)}. This is a standard metric for evaluating recommender systems, which is the average reciprocal rank for the items that a user actually interacts with. Higher MRR is better.

\item {\bf Mean Rank Percentile (MRP)}. We divide the rank of the ground truth interaction by the size of the candidate pool, and average over all test cases. This is useful for Pinterest dataset where the size of candidate pool vary over different interactions. Lower MRP is better.


\end{itemize}

\subsubsection{Baseline methods}
We compare HierTCN with a variety of baseline methods, which are summarized below.
To ensure a fair comparison, for all deep learning based model, we adjust the layer number and hidden units number such that all the models have very similar number of trainable parameters.

\xhdr{Rule-based models}
Rule-based models maintain a pool $\mathbf{P}_t$ that consists of the past $k$ interacted item embeddings $t$ for each user.
\begin{itemize}
    \item {\bf Moving average (MV)}. The moving average model outputs the average of all the item embeddings in $\mathbf{P}_t$.
    \item {\bf Maximum item similarity (MaxItem)} Rather than output a prediction, the MaxItem model can only evaluate a given candidate item embedding. An evaluation score is calculated as the highest dot product value between a candidate item embedding and embeddings in $\mathbf{P}_t$, which is then used for ranking the candidates.
\end{itemize}

\xhdr{Single-level sequence models}
Single-level sequence models take sequences of user interactions without being partitioned into sessions. To make a fair comparison, we include a session indicator, marking the start of a session, as a input feature. 
Specifically, we consider the following two sequence models:
\begin{itemize}
    \item {\bf TCN} We implement a TCN model with 6 blocks of convolutional layers, where each block consists of 2 temporal convolutional layers that have 128 filters with size 5. The dilation factor of each block is set to grow exponentially with the number of the block, which is 1,2,4,8,16,32 in our scenario. We also add residual connections \cite{he2016deep} between each block to facilitate training the model.
    \item {\bf GRU} We construct a GRU model by stacking 4 layers of GRU cells described in Equation \ref{eq:gru}, each with 200 dimensional hidden state.
\end{itemize}

\xhdr{Hierarchical sequence models}
We also compare HierTCN with the state-of-the-art hierarchical deep learning model, as well as a variant which we refer to as HierGRU.
\begin{itemize}
    \item {\bf HRNN} We implement a hierarchical GRU model following the paper \cite{quadrana2017personalizing}. Both high-level and low-level model are implemented with 4 layers of GRU cells, each with 128 dimensional hidden state. The high-level GRU's hidden state is used to initialized the low-level GRU, while the final hidden state of the low-level GRU is used to update the high-level GRU.
    \item {\bf HierGRU} We implement a baseline version of HierTCN, where the only difference is that we change the low-level model from TCN to GRU. Both the high-level GRU and the low-level GRU has 4 layers of GRU cells, each with 128 dimensional hidden state.
    \item {\bf HierTCN} This is our proposed model. The high-level model is a 4-layer GRU, each layer has 128 dimensional hidden state. The low-level model is a TCN with 4 blocks of convolutional layers, and the other settings are the same as a single-level TCN.
\end{itemize}


\begin{figure}[h]
    \centering
    \includegraphics[width=0.75\linewidth]{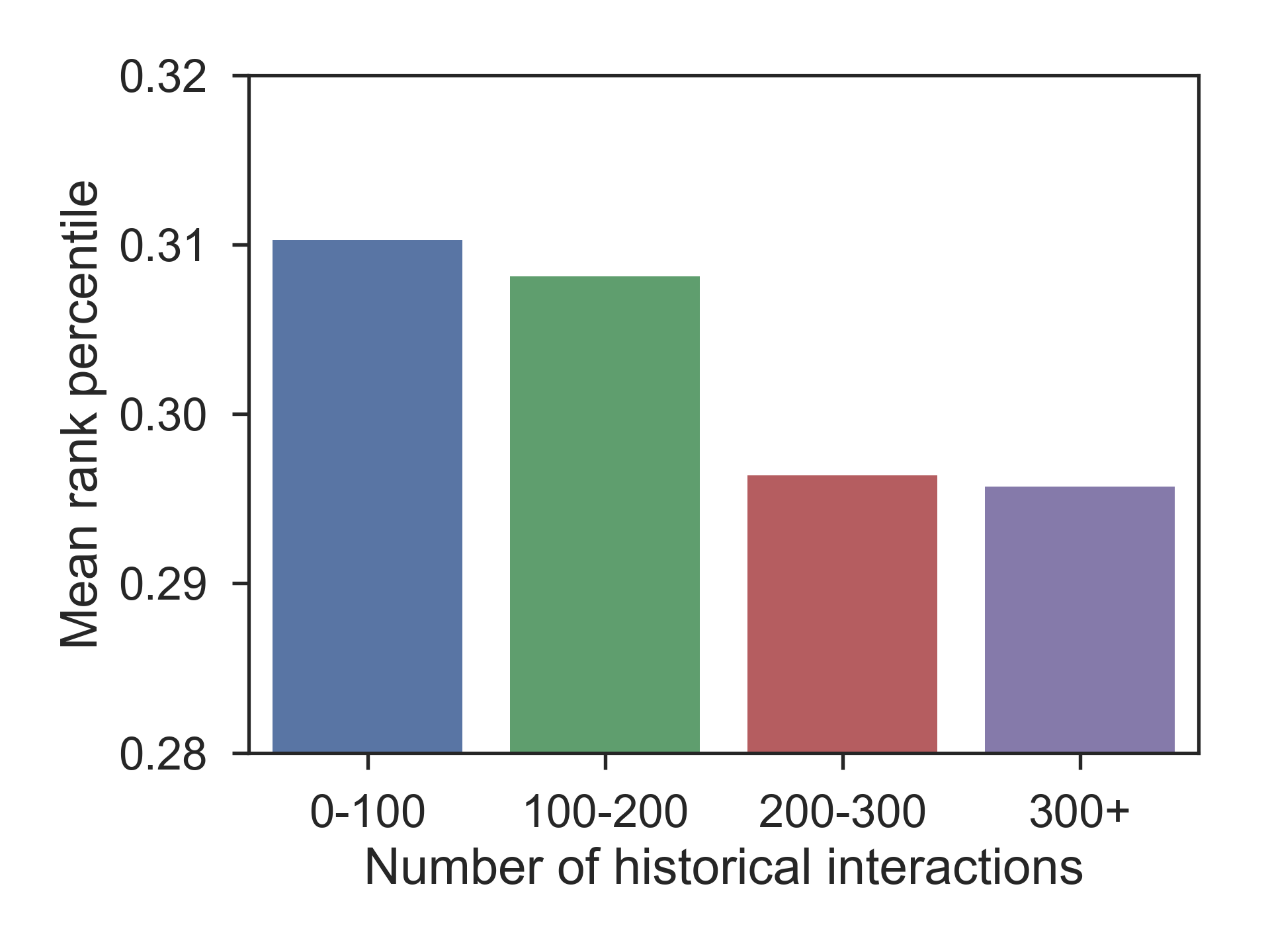}
    \vspace{-3mm}
    \caption{Summary of HierTCN's performance over users with different number of historical interactions. Lower is better.}
    \label{fig:my_label}
    \vspace{-3mm}
\end{figure}

\begin{figure}[h]
    \centering
    \includegraphics[width=0.75\linewidth]{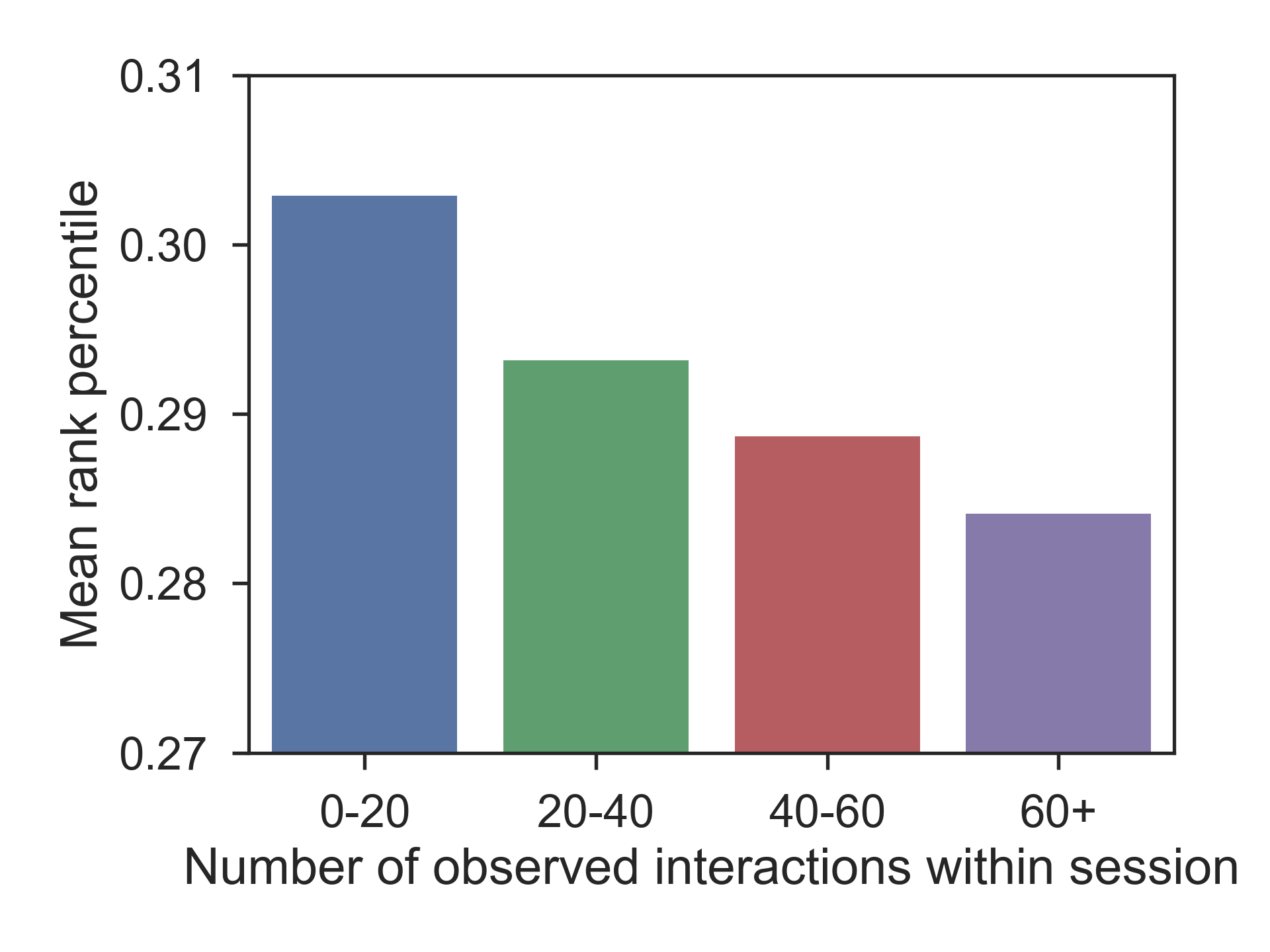}
    \vspace{-3mm}
    \caption{Summary of HierTCN's performance when observing different number of interactions within a session.}
    \label{fig:my_label2}
    \vspace{-3mm}
\end{figure}

\begin{figure}[h]
    \centering
    \includegraphics[width=0.75\linewidth]{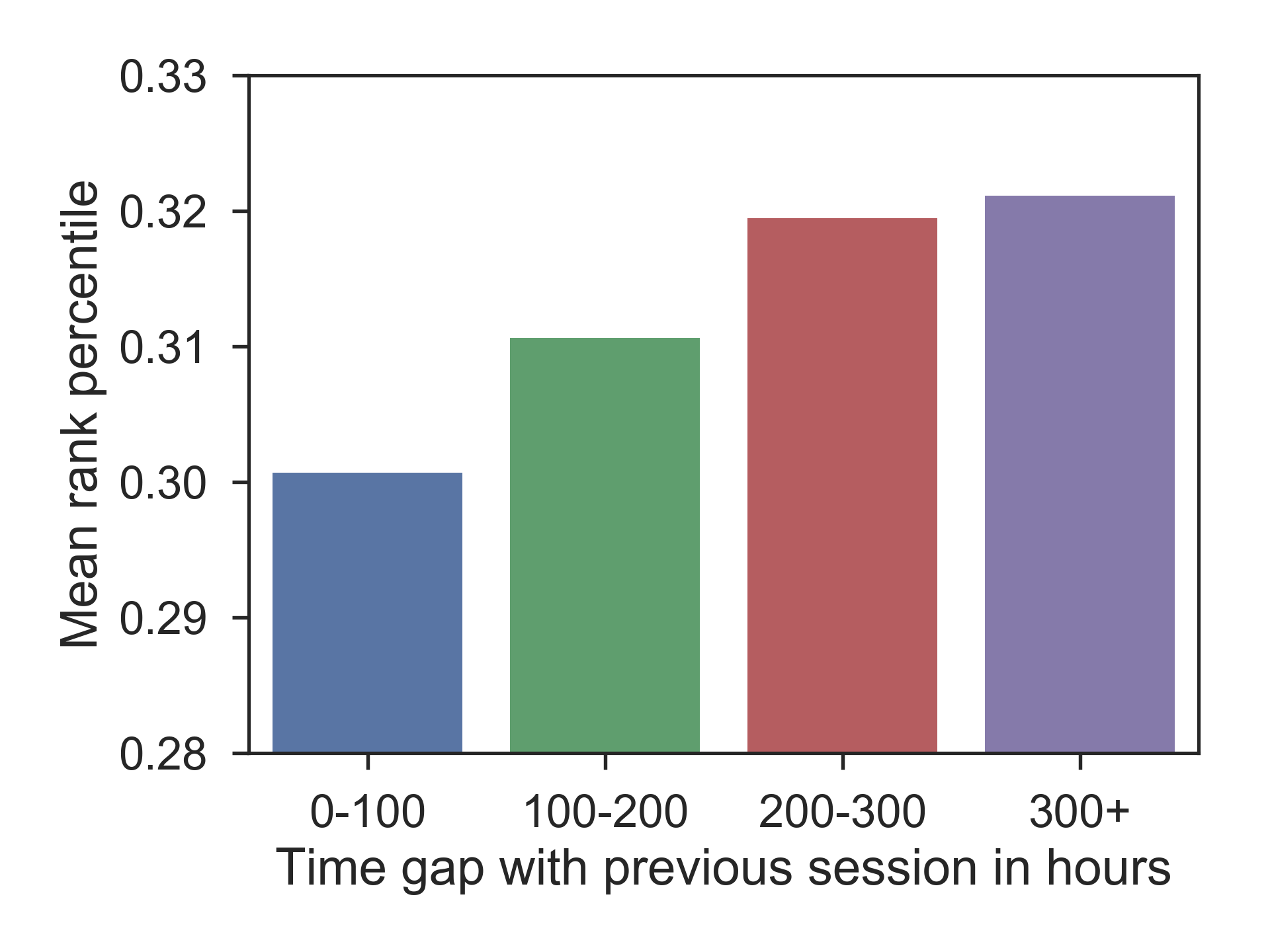}
    \vspace{-3mm}
    \caption{Summary of HierTCN's performance over sessions that have different time gap with previous sessions.}
    \label{fig:my_label3}
    \vspace{-3mm}
\end{figure}

\subsection{Experimental results}

\subsubsection{Results on Public XING dataset}

Table \ref{tb:cold} summarize the performance of all the models in XING dataset. Since items are represented as one-hot vectors, MV model cannot rank properly for unseen items thus only Recall@10 is reported. For the same reason, MaxItem is not implemented as the dot product between one-hot vectors is not meaningful.
Our model significantly outperforms all the baseline methods by a large margin, and achieves 30.4\% average performance improvement over the best baseline method (GRU). 
We observe that hierarchical GRU models converge very slowly, which may result from the complex gradient propagation path.
In this task, all the models directly predict the probability of each item over all possible items, resulting an output dimension of over 10 thousand. We train the models with the simplest cross entropy loss, and people can use a more complex loss function including \cite{quadrana2017personalizing} to achieve better performance.
The aim here is not to achieve the state-of-the-art performance on the XING dataset; rather, the results are used to show that the proposed HierTCN model can achieve more robust performance compared with baseline methods even in this simplified setting.








\begin{table}[h]
  \caption{The total time to finish one epoch training for the Pinterest dataset (in seconds).}
  \label{tb:time}
  \begin{tabular}{ccc}
    \toprule
  Dataset & Model & Training  \\
    \midrule
    Pinterest & \textbf{HierTCN} & \textbf{295.3s} \\
    & HierGRU & 747.8s \\
    & HRNN \cite{quadrana2017personalizing} & 693.1s \\
    & TCN & \textbf{254.0s} \\
    & GRU & 776.8s \\
  \bottomrule
\end{tabular}
\end{table}

\begin{table}
  \caption{Cold-start performance of HierTCN with different objective functions. 10\% of the training data are used for the experiment.}
  \vspace{-1mm}
  \label{tb:loss}
  \begin{tabular}{cccccccccc}
    \toprule
   Dataset & Metric & L2 & NCE & BPR & Hinge \\
    \midrule
        Pinterest & Recall@1 &  0.167&  0.192 & 0.183  & \textbf{0.201}  \\
    & Recall@5 & 0.611 &  0.642& 0.632&   \textbf{0.657} \\
    & Recall@10 & 0.817 &  0.838& 0.832&   \textbf{0.851}  \\
    & MRR & 0.360 & 0.386 &  0.377&   \textbf{0.396}  \\
    & MRP & 0.349 & 0.322 &  0.331&   \textbf{0.309}  \\
  \bottomrule
\end{tabular}
\vspace{-1mm}
\end{table}

\subsubsection{Results on Large-scale Pinterest dataset}

Table \ref{tb:cold} and Table \ref{tb:pinterest_warm} summarize the performance of all the models in the large-scale Pinterest dataset.
For cold-start evaluation, HierTCN outperforms the best hierarchical baseline by 18\% in Recall@1, 10\% in MRR.
The gain over the best single-level sequence model is 6\% in Recall@1, 3\% in MRR.
The gain over the best rule-based model is  24\% in Recall@1, 13\% in MRR.
From the results, we can see that deep learning approaches perform much better than rule-based models, since users are possible to interact with hundreds of millions of items in Pinterest, and thus rule-based models are incapable of capturing this complex dynamics.
Among deep learning approaches, it is interesting to see that hierarchical GRU models perform worse than single-level sequence models; empirically, we find that hierarchical GRU models converge slowly when we use the same learning rate as other models (0.001), while experiencing performance oscillation when we try to increase the learning rate.
On the contrary, HierTCN consistently outperforms the single-level sequence models, and we do not observe issues for optimizing the model. This indicates that the hierarchical structure does capture more aspects of user interests.
For warm-start evaluation, HierTCN also significantly outperforms the baseline models, with on average 12\% improvement in Recall@1 and 5\% improvement in MRR, and we can find similar observations from the results.


\subsubsection{Running time}

Table \ref{tb:time} summarizes the running time of different methods. We report the running time to finish one epoch training for all the deep learning models.
We observe that models that make use of TCN are roughly 2.5 times faster than GRU-based models, which supports the argument that TCN is generally much faster than GRU. Moreover, adding the hierarchical structure in HierTCN only slightly affects the computation speed, owing to the fact that we only update the high-level model of HierTCN after each session and we use an efficient mini-batch generator. The efficient mini-batch generator makes the training of hierarchical GRU models even slightly faster than the single-level GRU model.

\subsubsection{GPU memory consumption}
GPU memory consumption consists of the memory to store model parameters, which is the same for all models in our experiments, and the memory to store the input data. TCN model takes a huge amount of memory to store the input data, because it has to keep the entire historical sequence to make a prediction. In contrast, HierTCN only needs to store the input sequence at each session. The specific reduction of memory consumption varies with the number of sessions that a user has, and we observe up to 90\% reduction in the experiments.


\begin{table}
  \caption{Effects of adding different regularizations to HierTCN. BN and Drop stands for batch normalization and dropout.}
  \vspace{-1mm}
  \label{tb:reg}
  \begin{tabular}{cccccccccc}
    \toprule
   Dataset & Regularization & None & Drop & BN & Drop + BN \\
    \midrule
        XING & Recall@10 &  0.139&  0.129 & 0.145  & \textbf{0.147}  \\
            & MRR & 0.071 & 0.066 &  0.073&   \textbf{0.075}  \\
            & MRP & 0.121 & 0.120 &  0.109&   \textbf{0.107}  \\
  \bottomrule
\end{tabular}
\vspace{-1mm}
\end{table}





\subsection{Performance analysis}

\subsubsection{Choice of objective functions}
\label{sec:objective_result}
Table \ref{tb:loss} summarizes the results of using different loss functions for HierTCN on the Pinterest dataset. We conduct experiments in the cold-start setting and use only 10\% of all the training data for this objective function comparison, thus the performance scores are worse than Table \ref{tb:cold}.
We observe that using hinge loss provides about 5\% improvement of Recall@1 and 3\% improvement of MRR over the best competing loss function (NCE). 
In addition, using negative samples significantly improves the performance, and we observe 20\% improvement of Recall@1 and 10\% improvement of MRR over L2 loss.

\subsubsection{Effects of dropout and batch normalization}

To make fair comparison with baseline methods, we do not add dropout or batch normalization in the experiments.
We conduct further experiments to examine the effects of adding dropout and batch normalization.
From Table \ref{tb:reg}, we can see that adding dropout out alone does not improve the model performance, while adding batch normalization do help. When combining both techniques, there is further performance improvement. In addition, when adding batch normalization, we observe significant faster convergence; however, the model eventually overfits, and thus doing early stopping is necessary.

\subsubsection{Model behavior analysis}
We conduct further analyses to understand the performance of HierTCN under different scenarios. We select MRP as the metric and lower is better.
To examine the overall performance of HierTCN, we summarize the performance of HierTCN over users with different historical interactions in Figure~\ref{fig:my_label}. It is clearly shown that HierTCN's performance increases as more interactions are observed.
We analyze the performance of the low-level model by summarizing HierTCN's performance over different number of observed interactions within a session. Figure~\ref{fig:my_label2} shows that as more interactions are observed within a session, the performance of HierTCN gets better.
Finally, we examine the performance of the high-level model by summarizing the performance of sessions that have different time gaps with previous sessions. Figure~\ref{fig:my_label2} shows that as the session's time gap increases, a user's behavior becomes less predictable and the performance of HierTCN decreases. In summary, these analyses show that all the components of HierTCN perform reasonably under different scenarios.

\begin{figure}[t]
    \centering
    \vspace{-1mm}
    \includegraphics[width=\linewidth]{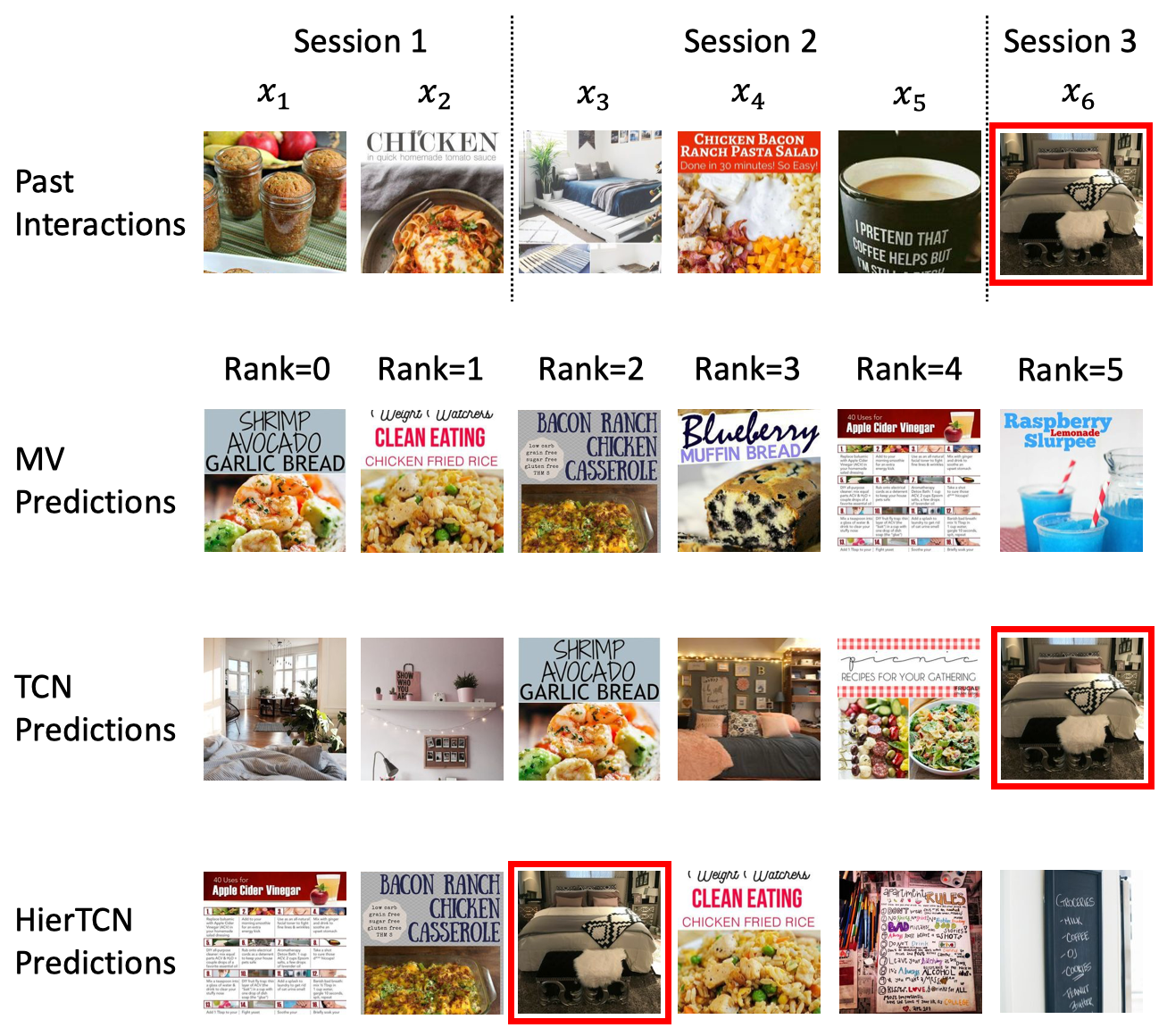}
    \vspace{-4mm}
    \caption{Recommendations given by different models. The first row shows the past interactions of a user, and the following 3 rows show the top-6 recommendations at time step 6 made by different models. The image marked with a red box is the ground-truth interaction.}
    \label{fig:html}
    \vspace{-4mm}
\end{figure}

\subsubsection{Visualization of recommendations}
Finally, we select a typical example to illustrate the superior performance of HierTCN.
The first row in Figure \ref{fig:html} shows the past interactions of a user, which includes items related to food and bed.
The following three rows illustrate the performance given by a rule-based model, a single-level sequence model and HierTCN.
From the results we can see that rule-based model recommends all food-related items because they are prevalent in the past interactions, TCN exaggerates the existence of bed-related item and recommends too many items related to furniture, and HierTCN reaches a balance between recommending both types of items and achieves better performance.

















\section{Conclusion}
In this paper, we have proposed Hierarchical Temporal Convolutional Networks for real-time large-scale recommender systems. By designing a novel hierarchical model using RNN and TCN, our model efficiently captures different levels of user interests and combines RNN's benefits of maintaining long-term hidden states and TCN's ability of conducting efficient and effective computation. We proposed a framework for large-scale dynamic recommender systems and applied HierTCN to a real-world dataset that contains millions of users and billions of activities. Compared with the state-of-the-art methods, HierTCN achieves superior performance and is much more scalable.

\section{Acknowledgement}
The authors thank Kaifeng Chen, Ruining He, and Fei Liu for their help on idea discussions and data preparation.



{
\bibliographystyle{ACM-Reference-Format}
\balance
\bibliography{bibli.bib}
}

\begin{appendix}
\end{appendix}

\end{document}